\documentclass[preprint,aps,prd,showpacs,nofootinbib]{revtex4}
\usepackage{epsfig}
\usepackage{graphicx}

\usepackage{epsfig}
\usepackage{graphicx,amsmath}

\newcommand\ba{\begin{eqnarray}}
\newcommand\ea{\end{eqnarray}}
\newcommand\nn{\nonumber}
\newcommand{\be}{\begin{equation}}
\newcommand{\ee}{\end{equation}}

\begin{document}

\title{General analysis of two--photon exchange in elastic electron--$^4\!He$
scattering and $e^++e^-\to \pi^++\pi^-.$ }

\author{G. I. Gakh }

\altaffiliation{Permanent address: \it  National Science Centre "Kharkov Institute of 
Physics and Technology", Akademicheskaya 1,  61108 Kharkov, Ukraine}

\email{gakh@kipt.kharkov.ua}

\affiliation{\it DSM/IRFU/SPhN, CEA/Saclay, 91191 Gif-sur-Yvette
Cedex, France}

\author{ E. Tomasi--Gustafsson}

\affiliation{\it DSM/IRFU/SPhN, CEA/Saclay, 91191 Gif-sur-Yvette
Cedex, France}
\date{\today}

\begin{abstract}
Using a general parametrization of the spin structure of the matrix element
for the elastic $e^-+^4\!He$ scattering and for the annihilation $e^++e^-\to \pi^++\pi^-$ reactions in terms of two complex amplitudes, we derive general properties 
of the observables in presence of two--photon exchange. We show that this mechanism induces a specific dependence of the differential cross section on the angle of the emitted particle. We reanalyze the existing experimental data on the differential cross section, for elastic electron scattering on $^4\!He$, in the light of this result.
\end{abstract}
\maketitle

\section{Introduction}

The measurement of the electromagnetic form factors (FFs) of hadrons and nuclei
in the space--like region of the momentum transfer squared has a long history (for a recent review see Ref. \cite{Pe07}).
The electric and magnetic FFs were determined both for the proton
and the neutron using two different techniques: the Rosenbluth separation
\cite{Ro50} and the polarization transfer method \cite{Re68}. It turned out that
the measurements of the ratio of the proton electric to magnetic FF using these two methods lead to different results, and the difference is increasing when $Q^2$ (the four--momentum transfer squared) increases \cite{Jo00,Ar04}. Possible explanations rely on the calculation of radiative corrections to the unpolarized cross section of elastic electron--nucleon
scattering: the necessity to include higher orders \cite{ETG07,By07} or to introduce a $2\gamma$ contribution  \cite{Twof}.

Hadron electromagnetic FFs are also investigated in the time--like region. Most of the data were obtained for the
$\pi$-- and $K$--mesons. The data on nucleon FFs in the time--like
region are scarce, and a precise separation of the electric and magnetic FFs has not been done yet. Recent data have been derived by the ISR method from the BABAR collaboration \cite{BABAR}.  Unexpected results have been
observed in the measurements of the nucleon FFs in the time--like
region (for a recent review see Ref. \cite{Ba06}). 

If the $2\gamma$ mechanism become sizable, the straightforward
extraction  of the FFs from the experimental data would be no longer possible \cite{Gu73}. It is known that double scattering dominates in
collisions of high--energy hadrons with deuterons at high $Q^2$ values
\cite{Fr73}, and in this paper it was predicted that $2\gamma$ exchange can represent a 10$\% $ effect in the elastic electron--deuteron scattering at $Q^2~\cong$ ~1.3 GeV$^2$ compared to the main ($1\gamma$) mechanism. At the same time the importance of the two--photon--exchange mechanism was considered in Ref. \cite{Bo73}. 

The reason that the $2\gamma$ mechanism, where the momentum transfer is equally shared between the two virtual photons, can become important with increasing $Q^2$ is that, due to the steep decrease of the FFs, such contribution can compensate the extra factor of $\alpha$ ($\alpha$=1/137 is the fine structure constant for the electromagnetic interaction). Perturbative QCD and quark counting rules \cite{Ma73,Le80} predict the dependence of FFs on the momentum transfer squared and, in particular, a steeper decreasing of FFs as the number of constituents particles involved in the reaction increases. Therefore, at the same value of $Q^2$, the relative role of two photon exchange with respect to the main mechanism, the one-photon exchange, is expected to be more important for heavier targets, as $d$, $^3\!He$, or $^4\!He$, than for the nucleon. The search of model independent evidence of two-photon exchange in the experimental data, which should appear as a non linearity of the Rosenbluth fit, was done in case of deuteron in Ref. \cite{Re99} and in case of proton in Ref. \cite{ETG05}. No evidence was found, in the limit of the precision of the data.

Let us note, that these experiments are sensitive to the real part of the interference between one and two photon exchange. A very precise measurement of the transverse beam spin asymmetry in elastic electron proton scattering is compatible with a non zero imaginary part of the two-photon exchange amplitude \cite{Ma04}. Recently, the HAPPEX
Collaboration at Jefferson Laboratory has measured the transverse beam spin 
asymmetry for elastic electron scattering from proton and $^4\!He$ target. It is the first measurement of the asymmetry from a nucleus and it appears to be non--negligible \cite{K07}. 
 
The $2\gamma$ contribution should also manifest itself in the time--like region. Theoretically this problem was firstly discussed in
Ref. \cite{Pu61} for the case of the annihilation of a $e^+e^-$--pair
into a $\pi\pi $ pair. The general analysis of the polarization phenomena
in the reaction $\bar p+p\to e^++e^-$ and in the time reversal channel, taking into account the $2\gamma$ contribution, was done in Ref. \cite{Ga06}. An analysis of the BABAR data does not show evidence of two photon contribution, in the limit of the uncertainty of the data \cite{ET07}.

In this paper we consider the problem of the two photon contribution in heavier targets. In case of spin 1/2 particles, as $^3\!He$ or $^3\!H$, one can apply the same model independent statements as for the nucleon \cite{Re04}. For spinless particles, the formalism will be derived in this work.

From the experimental point of view, the following reactions involving spinless particles, are easily accessible:
\be
e^-(p_1) +^4\! He(q_1)\to e^-(p_2)+^4\!He(q_2),
\label{eq:eq1}
\ee
and
\be
e^+ +e^-\to \pi^++\pi^-.
\label{eq:eq2}
\ee
The $^4\!He$ nucleus plays a special role among the few--body systems. It has
much higher density, comparable to the one of heavier nuclei. The effects
due to many--body forces and correlations are expected to be more important
than in the $A=3$ systems. Various models reproduce quite well the
$^4\!He$ binding energy and can be further tested by comparing the corresponding
FF to the data.

Data exist for the $^4\!He$ charge FF, and
the FF measurements extend to large $Q^2$ \cite{Si01}. The highest momentum transfers were achieved by an experiment carried out at
SLAC. The $^4\!He$ FF was measured up to $Q^2=64$ fm$^{-2}$ where the counting rate dropped to one event/week \cite{Ar78}. In this range the magnitude of the FF decreases by five orders of magnitude. In principle, at such values of $Q^2$, the $2\gamma$ contribution may appear from the data, 
since the Born (one--photon--exchange) contribution is expected to decrease faster than the $2\gamma$ one.

We derive here the expressions for the differential cross sections for the
case when the matrix element of the reactions (\ref{eq:eq1}) and (\ref{eq:eq2}) contains the
$2\gamma$ contribution. The parametrization of the
$2\gamma$ term is performed following a similar 
approach as used in the Refs. \cite{Re04}. We investigated also the effect
of non--zero lepton mass which may be relevant in the case of the muon scattering.

\section{The reactions $e^-+^4\!He\to e^-+^4\!He$ and $\pi +N\to \pi +N$}

We consider the $2\gamma$ contribution to elastic
electron--helium scattering, $e^-+^4\!He\to e^-+^4\!He$, in a model independent
way, based on general properties of the strong and the electromagnetic interactions. This approach is similar
to the one used for the analysis of the $2\gamma$ contribution to
the elastic electron--nucleon $e^-+N\rightarrow e^-+N$ \cite{Re04},
and to the proton--antiproton annihilation to the $e^+e^-$--pair \cite{Ga06}.

The spin structure of the matrix element for $e^-+^4\!He\to e^-+^4\!He$ can be established in
analogy with elastic pion--nucleon scattering \cite{Er88}, using the
general properties of the electron--hadron interaction, such as the Lorentz
invariance and P--invariance. In this respect, the discussion of the reactions $e^-+^4\!He\to e^-+^4\!He$ and $e^++e^-\to \pi^++\pi^- $ follows similar considerations, as the spin of the particle involved are the same and these  reactions are related by crossing symmetry. 

Taking into account the identity of the initial
and final states and the T--invariance of the strong interaction, the reactions
where a particle of the spin $0$ is
scattered by a particle with spin $1/2$, are described by two independent
amplitudes. So, the general form of the matrix element for the
$\pi +N\to \pi +N$ reaction as well as the $e^-+^4\!He\to e^-+^4\!He$ one, taking into account the $2\gamma$ contribution, are essentially identical and can be written as \cite{Er88}
\be
M(s,t)\simeq \bar u(p_2)\Bigg [A_1(s,t)+A_2(s,t)\hat Q\Bigg ]
u(p_1)\varphi (q_1)\varphi (q_2)^*,~ Q=\frac{1}{2}(q_1+q_2),
\label{eq:eq3}
\ee
$$Q=\frac{1}{2}(q_1+q_2), \ \  $$
where $\varphi (q_1)$, $\varphi (q_2)$ are the wave functions of the initial and final pions (or $^4\!He$), $u(p_1)$, $u(p_2)$ are the spinors describing the initial and final nucleon (or electron), and are functions of the corresponding four-momenta. Here $A_1$ and $A_2$ are two complex invariant 
functions of the variables $s=(q_1+p_1)^2$ and $t=(q_2-q_1)^2.$
At high energies, Feynman diagrams in QED are 
invariant under the chirality operation $u(p)\to \gamma_5 u(p)$, therefore, invariant structures in the matrix element which
change their sign under this transformation, such as $\bar u(p_2)u(p_1)$,  can be neglected in the unpolarized cross section, as they are small, proportional to the electron mass. However, such contributions are important in the analysis of the properties of some polarization observables and will be considered below.

Let us consider elastic
electron--helium scattering (including $2\gamma$ mechanism) and re-write Eq. 
(\ref{eq:eq3}) in the following general form:
\be
{\cal M}(s,q^2) =\frac{e^2}{Q^2}\bar u(p_2)\Bigg [mF_1(s,q^2)+
F_2(s,q^2)\hat P\Bigg ]u(p_1)\varphi (q_1)\varphi (q_2)^*=\frac{e^2}{Q^2}{\cal N},
\label{eq:eqmat}
\ee
where $\varphi (q_1)$
and $\varphi (q_2)$ are the wave functions of the initial and final helium, with $P=q_1+q_2$ and $u(p_1)$, $u(p_2)$ are the spinors of the initial and final electrons, respectively. Here $F_1$ and $F_2$ are two invariant amplitudes, which are,
generally, complex functions of two variables $s=(q_1+p_1)^2$ and
$q^2=(q_2-q_1)^2=-Q^2$ and $m$ is the electron mass.  The matrix element (\ref{eq:eqmat}) contains the helicity--flip amplitude $F_1$  which is proportional to the electron mass, which is
explicitly singled out. This small amplitude is not neglected here, because we will discuss a nonzero polarization observable, the 
single--spin asymmetry, which is proportional to $F_1$.

Therefore, two complex amplitudes, $F_i(s,q^2)$, $i=1,2$, fully describe the spin
structure of the matrix element for the reactions considered here, independently on the reaction mechanism, as the number of exchanged virtual photons.

In the Born (one--photon--exchange) approximation these amplitudes become:
\be
F_{1}^{Born}(s,q^2)=0, \  F_{2}^{Born}(s,q^2)=F(q^2),
\label{eq:eqampb}
\ee
where the function $F(q^2)$ is the helium electromagnetic charge form
factor depending only on the virtual photon four--momentum squared. Due to
the current hermiticity the FF $F(q^2)$ is a real function in
the region of the space--like momentum transfer.

The charge FF is normalized as $F(0)=Z,$ where $Z$ is the helium charge.

To separate the effects due to the Born and the 
two--photon exchange contributions, let us single out the dominant
contribution and define the following decompositions of the amplitude \cite{Re04}
\be
F_{2}(s,q^2)=F(q^2)+f(s,q^2).
\label{eq:eqtwo}
\ee
The order of magnitude of these quantities is $F_{1}(s,q^2)$ and
$f(s,q^2) \ \sim\alpha ,$ and $F(q^2) \ \sim \alpha^0$. Since the terms
$F_{1}$ and $f$ are small in comparison with the dominant one, we neglect
below the bilinear combinations of these small terms multiplied
by the factor $m^2$.

Then the differential cross section of the reaction (\ref{eq:eq1}) can be written as
follows in the laboratory (Lab) system: 
\be
\frac{d\sigma}{d\Omega}=\frac{\alpha ^2}{4M^2}\frac{E'^2}{E^2}
\frac{|{\cal N}|^2}{Q^4},
\label{eq:eqd}
\ee
where $Q^2=-q^2 $, $M$ is the helium mass, and $E(E')$ is the energy of the
initial (scattered) electron.

The differential cross section of the reaction (\ref{eq:eq1}), for the case of
unpolarized particles, has the following form in the Born approximation
\be
\frac{d\sigma_{un}^{Born}}{d\Omega}=\frac{\alpha ^2\cos^2\frac{\theta}{2}}
{4E^2\sin^4\frac{\theta}{2}}\Biggl [1+2\frac{E}{M}\sin^2\frac{\theta}{2}
\Biggr ]^{-1}F^2(q^2),
\label{eq:aqb}
\ee
where $\theta $ is the electron scattering angle in Lab system. Eq. (\ref{eq:aqb}) is consistent with the well known result for the differential cross section of the reaction (\ref{eq:eq1}), Ref. \cite{Ho56}. Including the $2\gamma$
contribution leads to three new terms
\ba
\frac{d\sigma_{un}}{d\Omega}&=&\frac{\alpha ^2\cos^2\frac{\theta}{2}}
{4E^2\sin^4\frac{\theta}{2}}\Biggl [1+2\frac{E}{M}\sin^2\frac{\theta}{2}
\Biggr ]^{-1}\Biggl\{F^2(q^2)+2F(q^2)Re~f(s,q^2)+|f(s,q^2)|^2+
\nn \\
&&+\frac{m^2}{M^2}\Biggl [\frac{M}{E}+(1+\frac{M}{E})\tan^2\frac{\theta}{2} 
\Biggr ]F(q^2)ReF_1(s,q^2)\Biggr\}. 
\label{eq:aqt}
\ea
Let us define the coordinate frame in Lab system of the reaction (\ref{eq:eq1}). The $z$ axis is directed along the momentum of the initial electron beam, the  $y$ axis is orthogonal to the reaction plane and directed along the vector
${\vec p}\times {\vec p}'$, where ${\vec p}({\vec p}')$ is the initial
(scattered) electron momentum, and the $x$ axis forms a left--handed
coordinate system.

Note that, in the case of the elastic scattering of 
transversally polarized electron beam on helium target, the $2\gamma$ contribution leads to a non--zero asymmetry, contrary to the Born
approximation.  This asymmetry arises from the 
interference between $1\gamma$ and $2\gamma$ exchange and can be written as: 
\be
A_y=\frac{\sigma^{\uparrow} -\sigma^{\downarrow}}
{\sigma^{\uparrow} +\sigma^{\downarrow}},
\ee
where $\sigma^{\uparrow} (\sigma^{\downarrow})$ is the cross section for electron beam
polarized parallel (antiparallel) to the normal of the scattering plane. This asymmetry is determined by the polarization component which is
perpendicular to the reaction plane: 
\be
A_y\sim {\vec s}_e\cdot \frac{{\vec p}\times {\vec p}'}
{|{\vec p}\times {\vec p}'|}\equiv s_y,
\ee
where ${\vec s}_e$ is the spin vector of the electron beam.
In terms of the amplitudes, it is expressed as:
\be
A_y=2\frac{m}{M}\tan\frac{\theta}{2}\frac{ImF_{1}(s,q^2)}{F(q^2)}.
\label{eq:eqasym}
\ee
Being a T--odd quantity, it is
completely determined by the $2\gamma$ contribution through the spin--flip amplitude $F_1(s,q^2)$ and, therefore, it is
proportional to the electron mass.

As one can see from Eq. (\ref{eq:aqt}), the extraction of the value of the helium form
factor from the measured cross section is determined by the real part of the
dominant two--photon exchange amplitude, if we neglect the small contributions due to the helicity--flip amplitude. On the
contrary, the $A_y$ asymmetry is determined only by the imaginary part of the
helicity--flip two--photon--exchange amplitude. Thus, the presence of this $A_y$ asymmetry must be taken into account in parity--violating experiments since it is a possible background in the measurement of the parity--violating asymmetry. Experimentally, for elastic  $e^-+^4\!He$ scattering,  a value of $A_y^{exp}(^4\!He)=-13.51\pm 1.34(stat)
\pm 0.37(syst)$ ppm  for $E =2.75$ GeV, 
$\theta = 6^0$, and $Q^2$=0.077  GeV$^2$ has been measured \cite{K07}, to be compared to a theoretical prediction
$A_y^{th}(^4\!He)\approx 10^{-10}$ which assumes that the target remains in its ground state
\cite{CH05}. This difference (by five orders of magnitude) was possibly  explained by a significant contribution of the excited states of the nucleus \cite{K07}.

Using the value of the measured asymmetry we can determine the size of the
imaginary part of the spin--flip amplitude $F_1$ for the experimental conditions
of the Jefferson Lab experiment \cite{K07}. From Eq. (\ref{eq:eqasym}) we  obtain $Im ~F_1\approx -F(q^2)$ for $\theta =6^0$. Assuming that $Re~F_1\approx Im~F_1$, then the contribution of the spin--flip amplitude to the differential 
cross section of the elastic electron--helium scattering is negligible due to
the small factor $m^2/M^2$. One may expect that the imaginary part of the 
non--spin--flip amplitude, namely, its two--photon--exchange part, is of the
same order as $Im~F_1$ since we singled out the small factor $m/M$ from the 
amplitude $F_1$. In this case we obtain an extremely large value for the 
two--photon--exchange mechanism, of the same order as the 
one--photon--exchange contribution itself, at such low $q^2$ value. Therefore, we can conclude that either our assumption, about the magnitudes of $Im~f$ and $Im~F_1$, is not correct, or the experimental results on the asymmetry are somewhat large.

\section{Reaction $e^++e^-\to \pi^++\pi^-$}

Let us consider the $2\gamma$ contribution to the reaction (\ref{eq:eq2}).
The matrix element of this reaction can be obtained from the expression
(\ref{eq:eqmat}) by the following substitution: $q_1\to -q_1,$ $p_2\to -p_2.$ As a result one has
\be
{\cal M}(q^2,t) =\frac{e^2}{q^2}\bar u(-p_2)\Bigg [mF_1(q^2,t)+
F_2(q^2,t)\hat R\Bigg ]u(p_1)\varphi (q_1)^*\varphi (q_2)^*=
\frac{e^2}{q^2}\bar {\cal N},
\label{eq:eqpi}
\ee
where $q=q_1+q_2,$ $R=q_2-q_1,$ $t=(p_1-q_1)^2$ and $q_1~(q_2)$ and
$p_1~(p_2)$ are the four--momenta of the final $\pi^- (\pi^+)$ meson and
electron (positron), respectively; $\varphi (q_1)$ and $\varphi (q_2)$ are
the wave functions of the final pions. Here $F_1$ and $F_2$ are two invariant
amplitudes, which are, generally, complex functions of two variables
$q^2$ and $t$.

In the Born (one--photon--exchange) approximation these amplitudes reduce to: 
\be
F_{1}^{Born}(q^2,t)=0, \  F_{2}^{Born}(q^2,t)=F(q^2),
\label{3}
\ee
where the function $F(q^2)$ is the pion electromagnetic charge FF
depending only upon the virtual photon four--momentum squared. In the region of  time--like momentum transfer due to the strong interaction in the final state the FF $F(q^2)$ is a complex function. The pion
FF has the following normalization: $F(0)=1$.

Again, to separate the effects due to the Born (one--photon exchange) and
$2\gamma$ contributions, let us single out the dominant
contribution and define the following decompositions of the amplitude
\be\label{5}
F_{2}(q^2, t)=F(q^2)+f(q^2,t).
\ee
The order of magnitude of these quantities is $F_{1}(q^2,t)$, 
$f(q^2,t)\sim\alpha$, and $F(q^2) \ \sim \alpha^0$. We again neglect
below the bilinear combinations of these small terms multiplied
by the factor $m^2$.

The differential cross section of the reaction (\ref{eq:eq2}) can be written as
follows in the center of mass system (CMS)
\be
\frac{d\sigma}{d\Omega}=\frac{\alpha ^2\beta}{8q^6}|\bar N|^2,
\label{eq:eq7}
\ee
where $\beta =\sqrt{1-4M^2/q^2}$ is the pion velocity in CMS and $M$ is the
pion mass.

The differential cross section of the reaction (\ref{eq:eq2}), for the case of
unpolarized particles, has the following form in the Born approximation
(neglecting the electron mass)
\be
\frac{d\sigma_{un}^{Born}}{d\Omega}=\frac{\alpha ^2\beta ^3}
{8q^2}\sin^2\theta |F(q^2)|^2,
\label{eq:eqtl}
\ee
where $\theta $ is the pion scattering angle in CMS. This expression
reproduce well known result for the differential cross section of the
reaction (\ref{eq:eq2}) \cite{G74}. The inclusion of the $2\gamma$
contributions leads to new terms:
\ba
\frac{d\sigma_{un}}{d\Omega}&=&\frac{\alpha ^2\beta ^3\sin^2\theta}
{8q^2}\Biggl\{|F(q^2)|^2+2ReF(q^2)f(q^2,t)^*+|f(q^2,t)|^2+\nn \\
&&+4\frac{m^2}{q^2}\frac{\cot\theta}{\sin\theta}
\Biggl [\cos\theta |F(q^2)|^2 +2ReF(q^2)(\cos\theta f(q^2,t)-\beta ^{-1}
F_1(q^2,t))^*\Biggr ]\Biggr\}. 
\label{eq:eqtl1}
\ea
Let us define the coordinate frame in CMS of the reaction (\ref{eq:eq2}). The
$z$ axis is directed along the momentum of the initial electron beam, $y$
axis is orthogonal to the reaction plane and directed along the vector
${\vec p}\times {\vec q}$, where ${\vec p}({\vec q})$ is the initial
electron (final pion) momentum, and the $x$ axis forms a left--handed
coordinate system.

Note that single--spin asymmetry for the reaction (\ref{eq:eq2}) is zero in the Born
approximation. But taking into account the $2\gamma$ contribution
leads to the non--zero asymmetry in the case of the scattering of the
transversally polarized electron beam. This asymmetry can be written as
\be
A_y=4\frac{m}{\sqrt{q^2}}\frac{1}{\beta}\frac{1}{\sin\theta}
\frac{ImF(q^2)F_1(q^2,t)^*}{|F(q^2)|^2},
\label{11}
\ee
and it is determined by the polarization component of the electron spin vector which is perpendicular to the reaction plane. Such asymmetry is a T-odd quantity, fully due to the $2\gamma$ contribution. It is determined by the spin--flip amplitude $F_1(q^2,t)$ and therefore it is proportional to the electron mass.

As it was shown in Ref. \cite{Re04}, symmetry properties of the amplitudes
with respect to the $\cos\theta \to -\cos\theta $ transformation can be
derived from the C invariance of the considered mechanism with the
$2\gamma$ contribution:
$$f(\cos\theta )=-f(-\cos\theta ),
~F_1(\cos\theta )=F_1(-\cos\theta ). $$
Let us consider the situation when the experimental apparatus does not
distinguish the charge of the pion. Then we measure the following sum of
the differential cross sections
\be
\frac{d\sigma_{+}}{d\Omega } =\frac{d\sigma}{d\Omega }(\cos\theta )+
\frac{d\sigma}{d\Omega }(-\cos\theta ).
\label{eq:eq18}
\ee
As shown in Ref. \cite{Pu61}, the sum of differential cross sections at $\theta$ and $\pi-\theta$ is not sensitive to the interference between the matrix elements corresponding to the one- and two--photon exchange diagrams, whereas the difference of the corresponding terms, is fully due to the presence of the $2\gamma$ contribution.

\section{Discussion of the experimental data}

\begin{figure}
\begin{center}
\includegraphics[width=12cm]{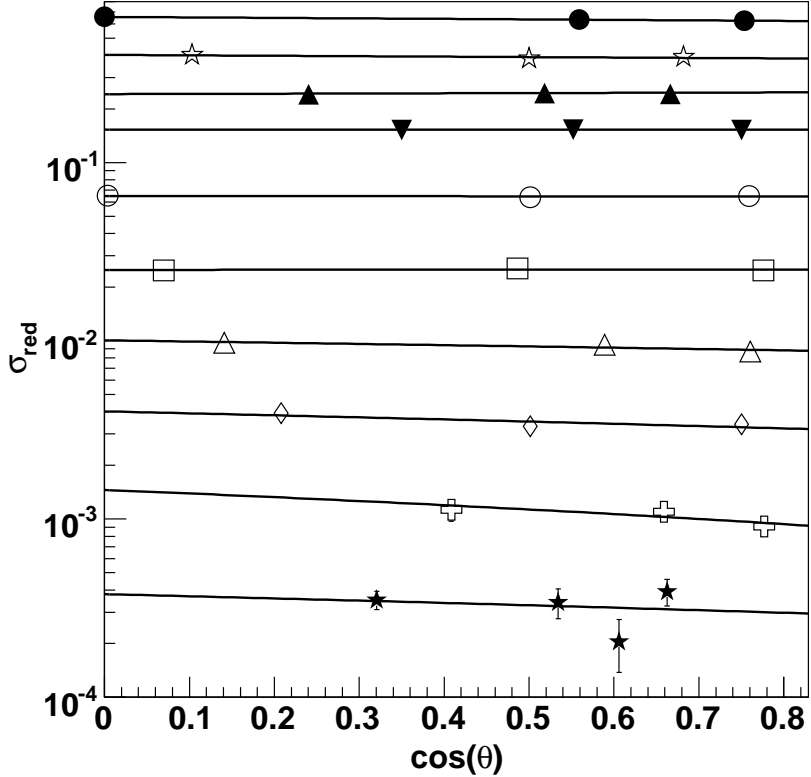}
\caption{\label{Fig:Ros} Reduced cross section as a function of $\cos\theta$, at $Q^2$=0.5, 1, 1.5, 2, 3, 4, 5, 6, 7, 8 fm$^{-2}$ (from top to bottom). The data are from Ref. \protect\cite{Fr67} and the lines are two parameter linear fits.  }
\end{center}
\end{figure}

In the Born approximation, the $^4\!He$ FF depends only on the momentum transfer squared, $Q^2$. The presence of a sizable $2\gamma$ contribution should appear as a deviation from a constant behavior of the cross section measured at different angles and at the same $Q^2$. In case of $^4\!He$ few data exist at the same $\bar Q^2$ value, for $Q^2<$ 8 fm$^{-2}$ \cite{Fr67}. We checked the deviation of these data from a constant value, with a two parameter fit, as a function of the cosine of the electron scattering angle: 
\be
\sigma_{red}|_{\bar Q^2}(\theta)= a +\alpha ~b\cos\theta .
\label{eq:eql}
\ee
No deviation from a constant is seen, the slope for each individual fit being always compatible with zero (see Fig. \ref{Fig:Ros}).  The results for the parameters of the individual fits and the $\chi^2$ are 
reported in Table I. 

\begin{table}[t]
\begin{tabular}{|l|l|l|l|}
\hline\hline
$ Q^2$ [fm $^{-2}$] & a $\pm \Delta a$  & b$\pm \Delta b$  & $\chi^2$\\
\hline
0.5&(66 $\pm$ 4) E-02	& -6 $\pm$  9	 &  0.1\\
1  &(0.40$\pm$ 3) E-02  & -3$\pm$ 8	 &  0.2\\
1.5&(0.24$\pm$ 2) E-02  & 1.0 $\pm$ 0.1	 &  0.1  \\
2  &(15 $\pm$ 2) E-03	& 0.0 $\pm$ 0.1	 &  0.1 \\
3  &(65 $\pm$ 4) E-03	& 0.$\pm$  1	 &  0.1\\
4  &(25 $\pm$ 2) E-03	& 0.0$\pm$ 0.4	 &  0.1 \\
5  &(101$\pm$ 8) E-04	& -0.2$\pm$ 0.2  &  0.5\\
6  &(40$\pm$ 5) E-04	&-0.1$\pm$ 0.1   &  0.6 \\
7  &(15$\pm$ 3) E-04	& -0.09$\pm$ 0.07&  1.0 \\
8  &(38$\pm$ 9) E-05	& -0.01$\pm$ 0.03&  1.0 \\
\hline\hline 
\end{tabular}
\caption{ For each Q$^2$ value, the intercept $a$ and the slope $b$  from the linear fit of the reduced cross section, as a function of $\cos\theta$, are given for the available $Q^2$ values (data from Ref. \protect\cite{Fr67}).}
\label{TableA}
\end{table}

From table \ref{TableA} a systematic negative sign for the slope appears, in particular at large $Q^2$, beyond the node of the FF. It could be the hint of a deviation from zero, which should increase at larger $Q^2$, if it results from a manifestation of $2\gamma$ exchange. 

To exploit the data points at higher $Q^2$, we did a two dimensional fit, in the variables $Q^2$ and $\cos\theta$:
\be
\sigma_{red}(Q^2,\cos\theta)=
(1-a^2Q^2)^6 e^{-b^2Q^2} [1+\alpha c_{2\gamma} Q^2\cos\theta].
\label{eq:eqfit}
\ee 
This parametrization is a simple modification of the one used in Ref. \cite{Ar78}, where the data were interpreted assuming $1\gamma$ approximation.
Eq. (\ref{eq:eqfit}) takes into account a $\cos\theta$ 
dependence as well as a smoother dependence on $Q^2$ which are both expected 
in presence of a $2\gamma$ contribution. Such parametrization gives a reasonable fit up to $Q^2$=20 $fm^{-2}$. The extracted FFs including (or not) the $2\gamma$ term in general overlap. A small difference can be seen at the node, which can be attributed to numerical instabilities, at large $Q^2$. Typically the additional $2\gamma$ term is lower that 1\%, excluding the factor of $\alpha$, and can be absorbed imposing an extra factor of $Q^2$ in the parametrization of the $2\gamma$ term. 

Data exist up to $Q^2\sim 40$ $fm^{-2}$, but two problems make difficult a reliable extraction of the $2\gamma$ contribution. From one side the $Q^2$ parametrization of FFs should have another functional dependence. In Ref. \cite{Ar78} a form as 
$\sigma_{red}(Q^2,\cos\theta)= a^2 e ^{-2bQ^2}, $
with $a=0.034\pm0.004$ and $b=2.72\pm 0.09$ was suggested.
On another side, the high $Q^2$ data were collected at the same scattering angle $\theta_e=8^0$, masking a possible angular sensitivity induced by the $2\gamma$ term. Therefore, from these data no extraction of a possible two photon contribution appears feasible.  

\section{Conclusions}
We have presented a parallel study of two reactions which involve 1/2 and zero spin particles: the elastic scattering 
$
e^- +^4\! He\to e^-+^4\!He$, and the 
$
e^+ +e^-\to \pi^++\pi^-
$ annihilation. We have derived general expressions for the matrix elements and for polarization phenomena, within a model independent formalism based on fundamental symmetries of the electromagnetic interaction and on crossing symmetry.

Let us summarize the main results of this paper. 

We have shown that the presence of $2\gamma$ exchange can be parametrized by two complex amplitudes instead of a real (complex) one for the scattering (annihilation) channel. The additional terms which appear in the cross section depend on the angle of the emitted particle, and 
should manifest 

- in the scattering channel, as an angular dependence of the reduced  differential cross section at fixed $Q^2$ 

- in the annihilation channel, as a charge asymmetry at the same emission angle, or in an asymmetric angular distribution of the emitted particle. 

The $2\gamma$ contribution could also be detected using a transversally polarized electron beam, which induces a T-odd asymmetry of the order of the electron mass.

An analysis of the existing data does not allow to reach evidence of the presence of the $2\gamma$ mechanism, as previously attempted for other reactions involving protons and deuterons. These conclusions hold including  Coulomb corrections, which are of the order of few thousandth.

The experimental data have been corrected from radiative corrections, using 
a method developed by Mo and Tsai \cite{Mo69}.  Radiative corrections modify the size and the angular dependence of the differential cross section and should be carefully taken into account. They depend on the experimental conditions, the kinematics and the acceptance, usually introduced as a cutoff in the energy spectra, which correspond to the maximal energy allowed for the emitted soft photons. Recently it appeared that higher order corrections should also be taken into account, as they are strongly dependent on the same kinematical variables which are relevant for the extraction of the FFs \cite{ETG07,By07}. In Ref. \cite{Fr67} it was mentioned that the applied corrections varied from 12\% to 24\%. Radiative corrections at higher order can be very large already at low $Q^2$, as the main effect is driven by $\log \frac{Q^2}{m^2}$ and they can be taken into account in a very effective way in frame of the structure fuinction approach \cite{Ku88}. A comparison between the first order calculations and the structure function approach will be published elsewhere.

In conclusion we stress the need for Rosenbluth experiments at larger $Q^2$ and for measurements of the single--spin asymmetry with a transversally polarized electron beam at other $Q^2$ values.

We also stress the need of taking into account radiative corrections at high order. The effect of large logarithm can appear already at low $Q^2$ values.

\section{Aknowledgment}
The formalism used in this work, was inspired by enlightning discussions with Prof. M. P. Rekalo. We thank Prof. E. A. Kuraev for interesting remarks concerning radiative corrections. This work is supported in part by grant INTAS Ref. No. 05-1000008-8328.


\end{document}